\documentclass[11pt]{article}
\usepackage{setspace}
\usepackage{amssymb}
\usepackage{amsmath}
\usepackage{amsfonts}
\usepackage{slashed}

\def\be{\begin{equation}}

\def\ee{\end{equation}}

\def\bea{\begin{eqnarray}}

\def\eea{\end{eqnarray}}

\makeatletter
\def\blfootnote{\xdef\@thefnmark{}\@footnotetext}
\makeatother

\begin{document}

\singlespace

\begin{flushright} BRX TH-???? \\
CALT-TH 2022-???
\end{flushright}

\vspace*{.3in}

\begin{center}

{\Large\bf  The Anthropic (and Mis-) Principle revisited, Steven Weinberg in memoriam }

{\large S.\ Deser}

{\it 
Walter Burke Institute for Theoretical Physics, \\
California Institute of Technology, Pasadena, CA 91125; \\
Physics Department,  Brandeis University, Waltham, MA 02454 \\
{\tt deser@brandeis.edu}
}

\end{center}

\begin{abstract}
 I review the role and meaning of the Anthropic Principle, particularly in its relevance
to particle physics as that subject evolves. This is neither a technical paper nor a resource letter,
but rather an easy introduction.
\end{abstract}

\section{Introduction}
Recent tributes to Steven Weinberg, our greatest living --- now alas gone --- theoretical physicist, have emphasized his use of this idea, among so many others. He made it a mainstream physics notion, worthy of discussing in a wider context, if only because of (despite?) its somewhat pompous title. Indeed, one might flippantly dismiss it on the grounds that, like the Holy Roman Empire's name, it is neither Anthropic nor a Principle, but rather a tautology. But despite this perhaps misleading fact, it is shorthand for a particular web of observational data, neither a panacea nor an act of desperation. To summarize, the Principle states the seemingly obvious requirement that the laws of Nature must permit the existence of some sort of sentient life, therefore our chemistry and biology: Like any experimental datum, this one puts stringent limits on our choice of (effective low energy, if not final) laws and their various constants such as masses, charges, Planck's and Newton's, as necessary conditions for organic chemistry, say, to be possible. Thus the word Anthropic is too narrow ---natural selection might well have evolved other forms of intelligence, as indeed it did until stumbling on us, and we might well not last long, either---but here we are and so must be accounted for! Also it provides a fascinating window on how our requirements change with time in the Wilsonian progression to higher energies. By the 1890s, organic chemistry was a flourishing autonomous discipline, but of course did not explain, but rather restated the Principle. Then the Bohr atom of 1915 gave it a microscopic basis; by 1926, Dirac could claim all of chemistry was subsumed by QM. The physics of, say, 1932, when the photon, electron and its neutrino, proton, neutron were in hand, as well as the nuclear models of the era could be said to provide a full basis. Indeed, Occam would be happy to stop here! But then came the modern standard model, which if anything, overshot the 1932 requirements, with its new particles, symmetries and confinement. Yet the principle is too tight in that it requires all the constants, such as masses, charges, etc., to be bounded, whereas we have no idea at all as to why they
are what they are! The above microscopic constraints are only some necessary conditions. There are also cosmological ones [1]; there too, Weinberg has contributed significantly [2], especially to tight limits on the value of the Cosmological Constant. The existence of sun-like stars with satellites of appropriate orbits is too far outside our scope to be useful, though stellar formation and stability are nontrivially relevant, while planets and their orbits are too ``accidental" to be of use. Indeed, new light has just now been shed [3] on our sun's origins and stability, a step forward. As for the word Principle, it is a bit too august --- all physics is based on Principle in the sense of agreement with the observed world. Descartes's famous cry was already proto-Anthropic, though alas only sufficient, not necessary --- as we know too well! In the late Fifties, Fred Hoyle and Willie Fowler explored the necessity of carbon energy levels (and iron's for the formation of the heavy elements). Note that I do NOT imply anything as silly as a pre-established harmony between our brains and Nature's laws.

\section{Details}
That was the short version of a complicated, and now enormously studied, subject, one worthy of wider intellectual appreciation. It is also at the heart of how science progresses. We look for data experimental or observational --- on which to base our models, as we modestly call our theories. This is not a tendentious process: it simply puts limits on their range. The Principle then is a name for one set of such data, namely and somewhat loosely, those that are required in order that our universe allows for our existence, which means more concretely two different sets: The first requires that the laws of gravity and matter make it possible for chemistry, in particular organic --- involving the element carbon --- to exist and evolve into biology via ever- more complicated molecular structures. The latter moves us into the second, very different arm, namely accidental, cosmological and galactic effects, such as the cosmological constant's value, formation and stability of stellar objects like our sun, ones that have its radiative properties and can live in that state for a long enough time before dying out or being involved in some galactic catastrophe, and even more fortuitously, to eject planets some of whose orbits permit the thermodynamic properties that make life possible, if not necessary. This latter is a very tall order indeed, as the vain search for exoplanets with these properties testifies. Not all necessary conditions can be fruitfully exploited! The main result here was to put fairly limits on the value of the cosmological constant.

 Fortunately, to borrow a sentence from the law, de minimis non curat physics: We are neither interested in, nor capable of calculating, such contingent events, as long as they are not manifestly forbidden by our models. So, being opportunistic is what the scientific method reduces to --- in this case finding any necessary conditions to explain what we observe, or discover upon experimentation guided by previous insights... and luck. The Principle's special
place is that it seemingly introduces a human element into the game. Yes, but --- really --- no more than limitations from entirely inhuman data! As a final thought, however, the above apparently trivial solar butterfly effect is why we Anthropos are so rare, if not unique, in our universe. I call it the Mis-Anthropic Principle: optimal planets and their orbits are forbiddingly unlikely to
occur! It no way invalidates the Principle's usefulness in bounding acceptable physical models; both are necessary.

\section{Summary}
Science proceeds using any tools it can; the Anthropic Principle is an example summarizing a web of observed
regularities, both cosmological and microscopic, that any physical models must
have room for, but in a way that evolves with theoretical progress, as we have illustrated.
Undoubtedly more such changes will appear in due time, perhaps an endless process!

\section*{Acknowledgements}
This work was supported by the U.S. Department of Energy, Office of Science, Office of High Energy Physics,  Award No. de-sc0011632.

\end{document}